\documentstyle[12pt,amsfonts]{article}
\newcommand{\define}{ \stackrel{\triangle}{=} }

\def\be{\begin{equation}}
\def\ee{\end{equation}}
\def\ba{\begin{array}}
\def\ea{\end{array}}

\def\d4{{\rm d}^4}

\begin{document}
%---------------------------------------------------
\vskip -1.0cm
\title{\bf Equation of Motion of a Spinning Test Particle
        in Gravitational Field}
\author{ {Ning WU}\thanks{email address: wuning@mail.ihep.ac.cn}
\\
\\
{\small Institute of High Energy Physics, P.O.Box 918-1,
Beijing 100049, P.R.China}
}
\maketitle
%\vskip 0.8in
%\noindent

%\vskip 0.8in
%\noindent

\begin{abstract}
Based on the coupling between the spin of a particle and
gravitoelectromagnetic field, the equation of motion of
a spinning test particle in gravitational field is deduced.
From this equation of motion, it is found that the motion
of a spinning particle deviates from the geodesic trajectory, and
this deviation originates from the coupling between the
spin of the particle and gravitoelectromagnetic field, which
is also the origin of Lense-Thirring effects.
In post-Newtonian approximations, this equation  gives out the
same results as those of Papapetrou equation.
Effect of the deviation of geodesic trajectory is detectable.

\end{abstract}


%-------------------------------------------------------

\newpage

\Roman{section}

\section{Introduction}

The motion of a test body in gravitational field is studied for a
long time. Through studying the motion of planets in solar system,
Newton found the famous Newton's law of universal gravitation\cite{01}.
In Newton's classical theory of gravity, the motion of a test particle
in gravitational field is driven  by the gravitational force on the
particle, and the corresponding equation of motion is
given by Newton's second law of motion.
In Einstein's general theory of relativity, a test body in gravitational
field always goes along the geodesic trajectory\cite{02,03}. Based on gauge
theory of gravity, it is  found that the geodesic equation is essentially
the same as the equation of motion given by Newton's second law of
motion\cite{04}. \\

In most textbook of General Relativity(GR)\cite{05,06}, only the motion
of a spinless mass point is discussed. The motion of mass point
with spin is rare discussed, for its equation of motion is not
consistent with the basic notion of the general relativity, especially
the cornerstone of the general relativity --- the equivalence principle.
However, the motion of a spinning test particles in general relativity
is widely discussed in literatures\cite{07,08,09,10}. In these discussions,
the equation of motion of a spinning test-particle is deduced from
the Bianchi identities, and the spinning test-particle is
treated as a pole-dipole
particle. It is known that, in classical mechanics, a point particle
can not have spin, for all angular momentum of an object originates
from circular motion of its inner composition, i.e., all angular momentum
in classical mechanics originates from orbital angular momentum.
Therefore, the spin of a test body in general relativity is defined by
\be \label{1.1}
S^{\alpha \beta} = \int
(\delta x^{\alpha} T^{\beta 4}
- \delta x^{\beta} T^{\alpha 4}) {\rm d}v,
\ee
where $T^{\alpha \beta}$ is the energy momentum tensor. The equation
of motion of a spinning test particle deduced in the
literatures \cite{07,08,09,10} can only applicable to the test
body whose spin originates from the orbital angular momentum
of its components. In quantum
mechanics, besides traditional orbital angular momentum, there exists
inner quantum spin, that is, a point particle can have non-vanish
spin quantum number in quantum mechanics. Because the inner quantum
spin of a point particle can not be defined by the equation (\ref{1.1}),
the equation of motion of a spinning test particle deduced in the
literature \cite{07,08,09,10} is not directly applicable to an elementary
particle with inner quantum spin. \\

In order to study the motion of an elementary particle with inner
quantum spin in gravitational field, a quantum theory of gravity
which is renormalizable is needed.
Quantum Gauge Theory of Gravity(QGTG) is first proposed in
2001\cite{11,12,13,14}. It is a quantum theory of gravity
proposed in the framework of quantum gauge field theory.
In 2003,  Quantum Gauge General Relativity(QGGR)
is proposed in the framework of QGTG\cite{15,16}.
Unlike Einstein's general theory of relativity, the cornerstone
of QGGR is the gauge principle, not the principle of equivalence,
which will cause far-reaching influence to the theory of gravity.
In QGGR, the field equation of gravitational gauge field is just
Einstein's field equation, so in classical level, we can set up
its geometrical formulation\cite{17}, and QGGR returns to
Einstein's general relativity in classical level.  The
field equation of gravitational gauge field in QGGR is the same as
Einstein's field equation in general relativity, so two equations
have the same solutions, though mathematical expressions of the
two equations are completely different. QGGR
is a perturbatively renormalizable
quantum theory, and based on it, quantum effects of
gravity\cite{18,19,20,21} and gravitational interactions of some basic
quantum fields \cite{22,23} can be explored. Unification of fundamental
interactions including gravity can be fulfilled in a simple and
beautiful way\cite{24,25,26,26a}. If we use the mass
generation mechanism which is proposed
in literature \cite{27,28}, we can propose a new
theory on gravity which contains massive graviton and
the introduction of massive graviton does not affect
the strict local gravitational gauge symmetry of the
action and does not affect the traditional long-range
gravitational force\cite{29}. The existence of massive graviton
will help us to understand the possible origin of dark matter.
In classical level, QGGR gives out the same theoretical
predictions as those of GR\cite{04}, and for non-relativistic
problems, QGGR can return to Newton's classical theory
of gravity\cite{30}. Based  on the coupling
between the spin of a particle and gravitoelectromagnetic
field, the equation of motion of spin can be obtained in QGGR.
In post Newtonian approximations, this equation of motion of
spin gives out the same results as those of GR\cite{31}.
\\

The transcendental foundations
of QGGR is quite different from those of
general relativity. Basic concept in QGGR
is that gravity is a kind of fundamental interactions in flat
Minkowski space-time, which is transmitted by gravitons. In other
words, in QGGR, space-time is always flat, and
geodesic curve is always a straight line. So, in QGGR,
geodesic equation can not be the equation of motion
of a mass point in gravitational field. In QGGR, the
motion of a test particle in gravitational field in classical level is
driven by the gravitational force on it, and the corresponding equation
of motion is given by the Newton's
second law of motion. In this paper, based on gauge principle and
Newton's second law of motion, the equation of motion of a spinning
test particle is deduced.
 \\

\section{Basics of Gauge Theory of Gravity}
\setcounter{equation}{0}

For the sake of integrity, we give a simple introduction to QGGR
and introduce some basic notations which is used in this paper. Details
on QGGR can be found in literatures \cite{11,12,13,14,15,16}. In gauge
theory of gravity, the most fundamental quantity is gravitational
gauge field $C_{\mu}(x)$, which is the gauge potential
corresponding to gravitational gauge symmetry. Gauge field
$C_{\mu}(x)$ is a vector in the corresponding Lie algebra, which
is called gravitational Lie
algebra. So $C_{\mu}(x) = C_{\mu}^{\alpha} (x)
\hat{P}_{\alpha} (\mu, \alpha = 0,1,2,3)$, where
$C_{\mu}^{\alpha}(x)$ is the component field and $\hat{P}_{\alpha}
= -i \frac{\partial}{\partial x^{\alpha}}$ is the  generator of
global gravitational gauge group. The gravitational gauge covariant
derivative is given by
\be \label{2.3}
D_{\mu} = \partial_{\mu} -
i g C_{\mu} (x) = G_{\mu}^{\alpha} \partial_{\alpha},
\ee
where $g$ is the gravitational coupling constant and matrix $G =
(G_{\mu}^{\alpha}) = ( \delta_{\mu}^{\alpha} - g C_{\mu}^{\alpha}
)$. Its inverse matrix is $G^{-1} = \frac{1}{I - gC} = (G^{-1
\mu}_{\alpha})$. Using matrix $G$ and $G^{-1}$, we can define two
important composite operators
\be \label{2.6}
g^{\alpha \beta} =
\eta^{\mu \nu} G^{\alpha}_{\mu} G^{\beta}_{\nu},
\ee
\be \label{2.7}
g_{\alpha \beta} = \eta_{\mu \nu} G_{\alpha}^{-1 \mu}
G_{\beta}^{-1 \nu},
\ee
which are widely used in QGGR. In QGGR,
space-time is always flat and space-time metric is always
Minkowski metric, so $g^{\alpha\beta}$ and $g_{\alpha\beta}$ are
no longer space-time metric. They are only two composite operators
which  consist of gravitational gauge field. The  field strength
of gravitational gauge field is defined by
\be \label{2.8}
F_{\mu\nu} (x) \define \frac{1}{-ig} \lbrack D_{\mu}~~,~~D_{\nu}
\rbrack = F_{\mu\nu}^{\alpha}(x) \cdot \hat{P}_{\alpha}
\ee
where
\be \label{2.12}
F_{\mu\nu}^{\alpha} = G_{\mu}^{\beta}
\partial_{\beta} C_{\nu}^{\alpha} -G_{\nu}^{\beta}
\partial_{\beta} C_{\mu}^{\alpha}.
\ee
\\

In QGGR, gravitational gauge field
$C_{\mu}^{\alpha}$ is a spin-2 tensor field. The field equation
of gravitational gauge field given by the least action principle
is just the Einstein's field equation\cite{15,16} \\

\section{Coupling between Spin and Gravitoelectromagnetic Field}
\setcounter{equation}{0}

In QED, we know that there is coupling between spin and magnetic field.
Transition of quantum states caused by this coupling gives out the famous
Landau energy level. In QGGR, similar phenomenon exists. In one of the
previous publications\cite{20}, the non-relativistic limit of the Dirac
equation in gravitational field and weak field approximations, an interesting
coupling between spin and gravitoelectromagnetic field is found, which has
the following forms
\be \label{3.1}
\Delta H = \frac{g}{2}
J^{\rho\sigma} F^0_{\rho\sigma},
\ee
with $J^{\rho\sigma}$ the spin of the particle. This coupling contributes
the following energy-momentum to that of the particle
\be \label{3.2}
\Delta p^{\gamma} = \frac{g}{2}
J^{\rho\sigma} F^{\gamma}_{\rho\sigma}.
\ee
In fact, the time component of $\Delta p^{\gamma}$ is just the
coupling energy $\Delta H$.
\\

There are two different kinds of forces originate from the coupling
between spin and gravitoelectromagnetic field. One force is given by
the space-time derivative of the potential energy
\be \label{3.3}
f_{1s \alpha} = - \partial_{\alpha} ( \Delta H )
= - \partial_{\alpha}
\left (\frac{g}{2} J^{\rho\sigma} F^0_{\rho\sigma} \right ).
\ee
Another force comes from the total time derivative of the coupling
energy-momentum $\Delta p^{\gamma}$ of the particle
\be \label{3.4}
f_{2s \alpha} = - \frac{\rm d}{{\rm d} \tau} ( \Delta p_{\alpha} )
= - \frac{\rm d}{{\rm d} \tau}
\left (\frac{g}{2}g_{\alpha\gamma} J^{\rho\sigma} F^{\gamma}_{\rho\sigma} \right ).
\ee
The total force originate from the coupling between spin and gravitoelectromagnetic
field is the sum of them
\be \label{3.5}
\ba{rcl}
f_{s \alpha} &=& f_{1s \alpha} + f_{2s \alpha}\\
&&\\
&= & - \partial_{\alpha}
\left (\frac{g}{2} J^{\rho\sigma} F^0_{\rho\sigma} \right )
- \frac{\rm d}{{\rm d} \tau}
\left (\frac{g}{2}g_{\alpha\gamma} J^{\rho\sigma}
 F^{\gamma}_{\rho\sigma} \right ).
\ea
\ee
\\

The expression in (\ref{3.5}) is neither Lorentz covariant nor gravitational
gauge covariant. We need to change its form to make it both Lorentz
covariant and gravitational gauge covariant, and the new form can return
to the equation (\ref{3.5}) in non-relativistic limit and weak gravitational
approximation. The above results are deduced
from the (\ref{3.1}) and (\ref{3.2}), which is the results of non-relativistic
limit of Dirac equation and weak gravitational approximation. In non-relativistic
limit, we have
\be \label{3.6}
\frac{{\rm d}x^0}{{\rm d} \tau} = \gamma  \simeq 1,
\ee
\be \label{3.7}
\frac{{\rm d} x^i}{{\rm d} \tau} = \gamma v^i \simeq 0,
\ee
and in weak gravitational approximation
\be \label{3.8}
g_{\beta \delta} \simeq \eta_{\beta \delta}.
\ee
After consider these approximations, the exact force which is Lorentz
covariant and gravitational gauge covariant should be
\be \label{3.9}
f_{s \alpha} = \left \lbrack
\partial_{\alpha} \left (
\frac{g}{2} g_{\beta \delta}
J^{\rho\sigma} F_{\rho\sigma}^{\delta} \right )
- \partial_{\beta} \left (
\frac{g}{2} g_{\alpha \delta}
J^{\rho\sigma} F_{\rho\sigma}^{\delta}
\right ) \right \rbrack
\frac{{\rm d} x^{\beta}}{{\rm d} \tau}.
\ee
The Lorentz covariance of (\ref{3.9}) is obvious, but its gravitational
gauge covariance is not obvious. In order to prove its gravitational
gauge covariance, we first introduce the following notation
\be \label{3.10}
A_{\alpha} = \frac{g}{2}
g_{\alpha \delta}
J^{\rho\sigma} F_{\rho\sigma}^{\delta}.
\ee
Then equation (\ref{3.9}) can be written in a simpler form
\be \label{3.11}
f_{s \alpha} = \left (
\partial_{\alpha} A_{\beta} - \partial_{\beta} A_{\alpha}
\right ) \frac{{\rm d} x^{\beta}}{{\rm d} \tau}.
\ee
\\

Define
\be \label{3.12}
\nabla_{\alpha} A_{\beta}
\define \partial_{\alpha} A_{\beta}
- \Gamma^{\gamma}_{\beta\alpha} A_{\gamma},
\ee
where $\Gamma_{\alpha\beta}^{\gamma}$ is defined by
\be \label{3.13}
\Gamma_{\alpha\beta}^{\gamma}
= \frac{1}{2} g^{\gamma\delta}
\left (
\frac{\partial g_{\alpha \delta}}{\partial x^{\beta}}
+ \frac{\partial g_{\beta \delta}}{\partial x^{\alpha}}
-\frac{\partial g_{\alpha \beta}}{\partial x^{\delta}}
\right ).
\ee
It can be proved that
\be \label{3.14}
\partial_{\alpha} A_{\beta} - \partial_{\beta} A_{\alpha}
= \nabla_{\alpha} A_{\beta} - \nabla_{\beta} A_{\alpha}.
\ee
Then equation (\ref{3.11}) can be changed into
\be \label{3.15}
f_{s \alpha} = \left (
\nabla_{\alpha} A_{\beta} - \nabla_{\beta} A_{\alpha}
\right ) \frac{{\rm d} x^{\beta}}{{\rm d} \tau}.
\ee
Or writing explicitly
\be \label{3.16}
f_{s \alpha} = \left \lbrack
\nabla_{\alpha} \left (
\frac{g}{2} g_{\beta \delta}
J^{\rho\sigma} F_{\rho\sigma}^{\delta} \right )
- \nabla_{\beta} \left (
\frac{g}{2} g_{\alpha \delta}
J^{\rho\sigma} F_{\rho\sigma}^{\delta}
\right ) \right \rbrack
\frac{{\rm d} x^{\beta}}{{\rm d} \tau}.
\ee
The above expression is obviously gravitational gauge covariant.\\

\section{Equation of Motion of a Particle with Spin }
\setcounter{equation}{0}

The equation of motion of spinless particle is discussed in the
literature \cite{04}. That equation of motion is deduced based on
the Newton's second law of motion. The basic concept in the deduction
is that gravity is treated as a kind of physical force, not space-time
geometry, and the motion of a particle in gravitational field is
driven by the gravitational force.
According to Newton's second law of motion, the equation of motion of
a particle with spin should be
\be \label{4.1}
\frac{{\rm d} P^{\mu}}{{\rm d}\tau}
= f^{\mu} + f^{\mu}_s,
\ee
where $P^{\mu}$ is the gravitational canonical energy-momentum
of the particle
\be \label{4.2}
P^{\mu} = G^{-1 \mu}_{\gamma} p^{\gamma},
\ee
$f^{\mu}$ is the Newtonian gravitational force whose form is obtained
in literature \cite{04}
\be \label{4.3}
f^{\mu} = g \eta^{\mu\nu} F_{\nu\lambda}^{\alpha_1}
T^{\lambda}_{g \alpha_1},
\ee
and $f^{\mu}_s$ is the force originating from the coupling between
spin and gravitoelectromagnetic field
\be \label{4.4}
f_s^{\mu} = G^{-1 \mu}_{\gamma} g^{\gamma \alpha} f_{s \alpha}.
\ee
In equation (\ref{4.3}), $T^{\lambda}_{g \alpha_1}$ is the gravitational
energy-momentum tensor of the particle
\be \label{4.5}
T^{\lambda}_{g \alpha_1}
= \frac{1}{2}
\left (g_{\alpha_1 \beta} G^{-1 \lambda}_{\alpha}
+ g_{\alpha_1 \alpha} G^{-1 \lambda}_{\beta}\right )
p^{\alpha} \frac{{\rm d} x^{\beta}}{{\rm d} \tau}.
\ee
Using all above relations, the equation of motion of a spinning particle
in gravitational field can be explicitly written out
\be \label{4.6}
\ba{rcl}
\frac{\rm d}{{\rm d} \tau}
\left ( G^{-1 \mu}_{\gamma_1} p^{\gamma_1}
\right )
&=& \frac{g}{2} \eta^{\mu\nu}
\left ( g_{\alpha_1 \beta} G^{-1 \lambda}_{\alpha}
+ g_{\alpha_1 \alpha} G^{-1 \lambda}_{\beta}
\right )
  F^{\alpha_1}_{\nu\lambda}
p^{\alpha} \frac{{\rm d} x^{\beta}}{{\rm d} \tau}
\\
&&\\
&&+ G^{-1 \mu }_{\gamma_1} g^{\gamma_1 \alpha}
\left \lbrack
\nabla_{\alpha} \left ( \frac{g}{2} g_{\beta \delta}
J^{\rho\sigma} F^{\delta}_{\rho \sigma}
\right )
- \nabla_{\beta} \left ( \frac{g}{2} g_{\alpha \delta}
J^{\rho\sigma} F^{\delta}_{\rho \sigma}
\right )
\right \rbrack \frac{{\rm d} x^{\beta}}{{\rm d} \tau}
\ea
\ee
Multiplying both side of the above equation with $G_{\mu}^{\gamma}$ and
summing over the index $\mu$, we get
\be \label{4.7}
\ba{rcl}
\frac{{\rm d} p^{\gamma}}{{\rm d} \tau} &+&
\left \lbrack  \frac{1}{2} G_{\mu}^{\gamma}
(\partial_{\alpha} G^{-1 \mu}_{\beta}
+ \partial_{\beta} G^{-1 \mu}_{\alpha})
- \frac{g}{2} \eta^{\mu\nu} G_{\mu}^{\gamma}
\left ( g_{\alpha_1 \beta} G^{-1 \lambda}_{\alpha}
+ g_{\alpha_1 \alpha} G^{-1 \lambda}_{\beta}
\right )
  F^{\alpha_1}_{\nu\lambda} \right \rbrack
p^{\alpha} \frac{{\rm d} x^{\beta}}{{\rm d} \tau} \\
&&\\
&=&  g^{\gamma \alpha}
\left \lbrack
\nabla_{\alpha} \left ( \frac{g}{2} g_{\beta \delta}
J^{\rho\sigma} F^{\delta}_{\rho \sigma}
\right )
- \nabla_{\beta} \left ( \frac{g}{2} g_{\alpha \delta}
J^{\rho\sigma} F^{\delta}_{\rho \sigma}
\right )
\right \rbrack \frac{{\rm d} x^{\beta}}{{\rm d} \tau}
\ea
\ee
Substitute equation (\ref{2.6}) and (\ref{2.7}) into equation
(\ref{3.13}), we get
\be \label{4.8}
\Gamma^{\gamma}_{\alpha\beta}
= \frac{1}{2} G_{\mu}^{\gamma}
(\partial_{\alpha} G^{-1 \mu}_{\beta}
+ \partial_{\beta} G^{-1 \mu}_{\alpha})
- \frac{g}{2} \eta^{\mu\nu} G_{\mu}^{\gamma}
\left ( g_{\alpha_1 \beta} G^{-1 \lambda}_{\alpha}
+ g_{\alpha_1 \alpha} G^{-1 \lambda}_{\beta}
\right )
  F^{\alpha_1}_{\nu\lambda}.
\ee
Then equation (\ref{4.7}) can be written into the following
simpler form
\be \label{4.9}
\frac{D p^{\gamma}}{D \tau}
= g^{\gamma \alpha}
\left \lbrack
\nabla_{\alpha} \left ( \frac{g}{2} g_{\beta \delta}
J^{\rho\sigma} F^{\delta}_{\rho \sigma}
\right )
- \nabla_{\beta} \left ( \frac{g}{2} g_{\alpha \delta}
J^{\rho\sigma} F^{\delta}_{\rho \sigma}
\right )
\right \rbrack \frac{{\rm d} x^{\beta}}{{\rm d} \tau},
\ee
where
\be \label{4.10}
\frac{D p^{\gamma}}{D \tau}
= \frac{{\rm d} p^{\gamma}}{{\rm d} \tau}
+ \Gamma^{\gamma}_{\alpha\beta}
p^{\alpha} \frac{{\rm d} x^{\beta}}{{\rm d} \tau}.
\ee
It can be easily proved that
\be \label{4.11}
\nabla_{\alpha} g_{\beta \gamma} =0,
\ee
\be \label{4.12}
\nabla_{\alpha} g^{\beta \gamma} =0,
\ee
then equation (\ref{4.9}) can be changed into the following form
\be \label{4.13}
\frac{D }{D \tau}
\left ( p^{\gamma}
+ \frac{g}{2} J^{\rho\sigma} F^{\gamma}_{\rho \sigma}
\right )
= \frac{g}{2} g^{\gamma \alpha} g_{\beta \delta}
\nabla_{\alpha} \left (
J^{\rho\sigma} F^{\delta}_{\rho \sigma}
\right )
 \frac{{\rm d} x^{\beta}}{{\rm d} \tau}.
\ee
This is the equation of motion of a spinning particle.
In the above equation, the term
$\frac{g}{2} J^{\rho\sigma} F^{\gamma}_{\rho \sigma}$
in the left hand side is just the $\Delta p^{\gamma}$
in equation (\ref{3.2}), which can be understood as the
interactive energy-momentum of the particle originate from
the coupling between spin and gravitoelectromagnetic field.
And
\be \label{4.13a}
p_t^{\gamma} =p^{\gamma}
+ \frac{g}{2} J^{\rho\sigma} F^{\gamma}_{\rho \sigma}
\ee
can be understood as the total energy-momentum of the
spinning particle.
The term in the right hand side can be explained as the
interactive force originate from the coupling between spin
and gravitoelectromagnetic field. \\

For the spinless particle, it's spin  $J^{\rho\sigma}$ vanishes
\be \label{4.14}
J^{\rho \sigma} = 0.
\ee
then equation (\ref{4.13}) becomes
\be \label{4.15}
\frac{D p^{\gamma}}{D \tau}
= 0,
\ee
which is the same as the equation of motion deduced in the literature
\cite{04}. This equation is just the traditional geodesic equation
in general relativity.\\

\section{Summary and Discussions}

In this paper, based on the coupling between spin and gravitoelectromagnetic
field, the equation of motion of a spinning particle in gravitational field
is deduced. This equation can be used to calculate the orbit of a mass
point with spin or gyroscope.
Contrary to general relativity, the basic concept in QGGR
is that, gravity is a kind of physical force on particles, and space-time
is always flat. For a spinning  particle, it not only feel the traditional
Newtonian gravitational force,  but also feel a new kind of gravitational
force which originates from the coupling between spin and gravitoelectromagnetic
field. This new kind of gravitational force depends on the spin of the
particle, which violates weak equivalence principle\cite{32}, for particles with
spin does not go along the geodesic curve. Because the spin of the particle
is a quantity of inner structure of the particle, not a quantity of space-time,
the fact that the orbit of a particle depends on its spin means that its orbit
is not completely determined by space-time geometry.
\\

In this paper, we have discussed the influence of the coupling between spin
and gravitoelectromagnetic field to the motion of a particle. Another important
influence of this coupling is that it causes precession of a gyroscope which
is discussed in literature \cite{31}. This effect had already been detected
by experiments\cite{33,34}. Results of the experiments on the precession of
a gyroscope show that the coupling between spin and gravitoelectromagnetic field
does exist. \\

In general relativity, the equation of a pole-dipole particle is given by
Papapetrou equation\cite{09,10}. Papapetrou equation can applied to the
spinning object, whose spin comes from the orbital angular momentum of
its constituents. In principle, Papapetrou equation can not be used to
discuss the motion of a particle with quantum spin, for quantum spin
can not be defined by equation (\ref{1.1}). It is found that, in
post-Newtonian approximations, equation (\ref{4.13}) gives out the
same results as those of Papapetrou equation. \\

Further study shows that, in some circumstances,  the effect of the
deviation of the geodesic curve is detectable, which will be discussed
in another paper\cite{35}.
\\

\end{document}